# A New Class of Boron Nanotube

Jing Wang, Ying Liu,* and You-Cheng Li

The configurations, stability and electronic structures of a new class of boron sheet and related boron nanotubes have been predicted within the framework of density functional theory. This boron sheet is sparser than those of recent proposals. Our theoretic results show that the stable boron sheet remains flat and is metallic. There are bands similar to the π-bands in graphite near the Fermi level. Stable nanotubes with various diameters and chiral vectors can be rolled from the sheet. Within our study, only the thin (8, 0) nanotube is semiconducting with a band gap of 0.44 eV, while all the other thick boron nanotubes are metallic independent of their chirality. It indicates the possibility, in the design of the nanodevice, to control the electronic transport properties of the boron nanotube through the diameter.

## Introduction

Since the discovery of $C_{60}$,[1] carbon nanotubes (CNT) as well as graphene, the precursor of the carbon fullerenes and CNTs, these molecules have attracted wide attention due to their novel electronic properties and great potential of applications.[2,3] Boron as carbon's neighbor to the left in the periodic table possesses a richness of chemistry second only to carbon. There is a growing interest in exploring the structure and energetics of pure boron clusters,[4-8] boron tubes with widths on the nanometer scale,[9-14] and boron containing molecules[15] because they are expected to have wide applications. It has been proposed that the most stable structure of $B_{20}$ is a double-ring tubular structure, which can be considered as the embryo of the single-walled boron nanotubes (BNTs).[5] The BNT was originally predicted by I. Boustani and A. Quandt.[9] In 2004, Ciuparu et al.[10] successfully synthesized pure boron single-walled nanotubular structures with diameter in the range of 3 nm and thus confirmed the suggested existence of BNTs. Recently, a stable hollow cage of $B_{80}$,[16] has been predicted based on ab initio calculations. Similar to the well-known carbon fullerene, this recent prediction of the boron buckyball $B_{80}$ cage has stimulated intensive studies on boron sheets (BS) and BNTs. Using density functional theory (DFT) within the ab initio supercell plane-wave pseudopotential total energy approach, a new class of flat stable boron sheet, the α-B sheet, was predicted to be the most stable boron sheet studied so far with more favorable energy than the buckled triangular sheet.[17] In addition, theoretical studies have shown that a series of BNTs may be obtained by rolling up the stable α-B sheet. From an electronic point of view, it might be expected that BNTs should always be metallic, independent of their structure, in contrast to CNTs, which are either semiconducting or metallic, depending on their diameter and chirality. However, first-principles calculations have indicated that the BNTs with small diameters open the band gap and are semiconductors.[18,19] Very recently, the electronic structure of α-tetragonal boron nanobelts was studied for the first time by electron energy-loss spectroscopy (EELS) and soft X-ray emission spectroscopy (SXES) and the results indicate that the boron nanobelts are either a semimetal or narrow-gap semiconductor.[20]

In this work, a new class of stable boron sheet has been constructed. Its binding energy per atom is shown to be close to that of the recently found stable α-B sheet[17,18]. This suggests the similar possibility of this new BS in sample preparation to α-B sheet. The stability and electronic properties of the new sheet and the corresponding BNTs rolled from it, were investigated within the framework of DFT.

## Results and Discussion

Figure 1 gives the structure of the new class of boron sheet, which is sparser than the α-B sheet[17,18]. After relaxation of the supercells, the boron sheet remains flat and is metallic. The binding energy per atom of this boron sheet, defined by $E_b=[E(BS)-nE(B)]/n$, is 5.93 eV and it is very close to that of the α-B sheet, 5.97 eV. The band structure, as well as the density of state (DOS), is also shown in Fig. 1. There are bands similar to the π bands in the graphene. Figure 2 gives the charge density of the boron sheet. The presence of significant charge density at the Fermi level indicates that the boron sheet is metallic.

The boron sheet can be rolled into boron nanotubes and the edges stitched seamlessly together. As shown in Fig. 1 (a), there list two kinds of vectors: the primary vectors of the hexagonal lattice ($a_1$, $a_2$) used for CNT and the primary vectors of BS lattice ($b_1$, $b_2$). Furthermore, the reduced symmetry of the BS makes the ($b_1$, $b_2$) based on a rectangular lattice more appropriate to classify the BNTs than using the hexagonal vector. That would be more convenient to label boron nanotubes with integer multiples of the basis vectors.[14] For the chiral vector ($p$, $q$) of the boron sheet lattice, it corresponds to the vector ($n$, $m$) of the hexagonal lattice with $p=2n+m$ and $q=3m/5$.

Calculations were performed on these tubes with different diameters in the range 3.16-14.58Å and various chiral vectors ($p$, $q$), in order to explore their stabilities and electronic properties. The binding energies per atom of the considered BNTs are shown in Fig. 3. It can be seen that the binding energy is approaching to that of the BS as the diameter increases. The BNT with large diameter has relatively high stability and the stability is increased with the increase of the diameter. The electronic structure of the (*p*, 0) (*p*=8,10,...,18), (0, *q*) (*q*=4, 6, 8), and (*p*, *q*) tubes were calculated. All the calculated results indicate that the BNTs rolled from this boron sheet are metals, except for the (8, 0) thin nanotube, for which there is an energy gap of 0.44 eV. It is noted that in the hexagonal boron ring of the (8, 0) thin nanotube there is a next-nearest neighbor B-B bond, which makes the density of the thin boron tube closer to that of bulk boron[21,22] and indicate the semiconducting properties. Some selected typical structures and their corresponding band structures and DOSs are shown in Fig. 4. For the band structures of the metallic boron nanotubes, there are several bands in the vicinity of the Fermi level, which ensures a large carrier density. Take the (0, 8) tube for example, the bands are degenerate at the *Γ*-point and these bands are highly dispersive along the *Γ-X* axis, implying that the effective mass of the charge carriers should be very small, leading to high mobility and high conductivity. Hence the BNTs could have potential as the metallic interconnects in electronic devices.

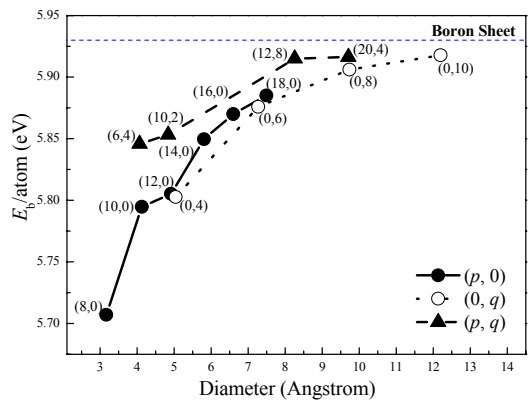

Figure 3. The binding energy per atom ($E_b$/atom) of the boron nanotubes *vs* the diameters.

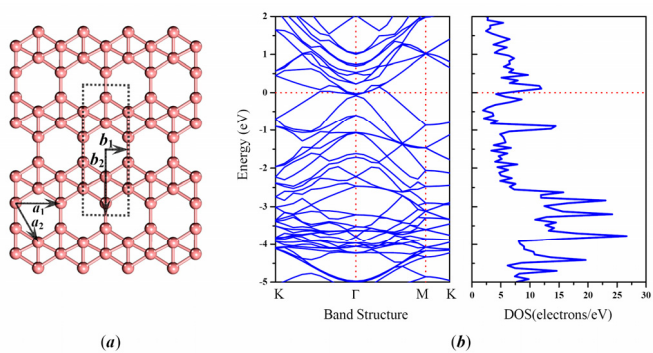

Figure 1. (*a*) The geometric structure; (*b*) the band structure and density of states (DOS) for the boron sheet.

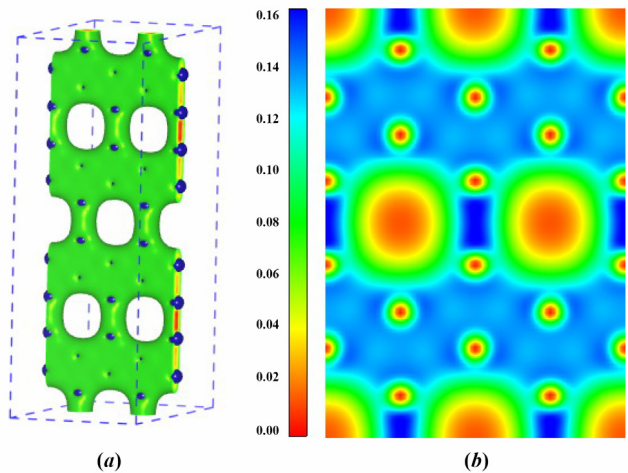

Figure 2. The charge density of the boron sheet. (*a*) Solid image with iso-value set at 0.08 electrons/a.u.³; (*b*) the (0 0 1) plane of the solid image.

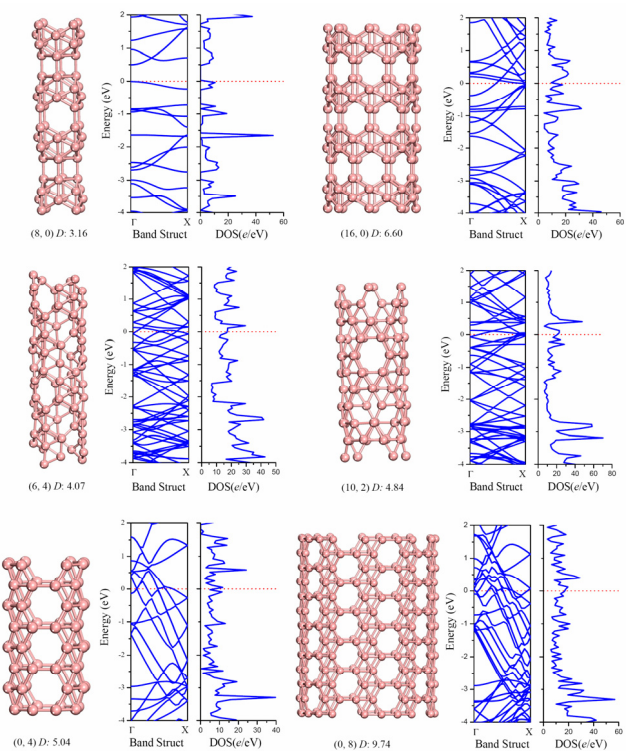

Figure 4. The geometric structure, band structure and density of state (DOS) of the boron nanotubes. Below each BNT give the chiral vector (*p*, *q*) and the diameter (*D*) in unit of Å.

## Conclusion

In summary, the structural and electronic properties of a novel boron sheet and the related boron nanotubes have been investigated at the DFT-GGA level. The new class of boron sheet is sparser than the *α*-B sheet and after relaxation it remains flat and metallic. Within the scope of our research, except for the (8, 0) thin tube, all the nanotubes rolled from this sheet are metallic independent of their chirality. Our predicted boron sheet and boron nanotubes have similar electronic properties, which may make them good candidates for fabrication.

## Theoretical Methods

The present calculations of boron sheets and boron nanotubes were performed by using the Vienna *ab initio* simulation package (VASP).[23] The approach is based on the DFT in a plane-wave basis set with Vanderbilt ultrasoft pseudopotentials.[24] The generalized gradient approximation (GGA) given by Perdew-Wang (PW91)[25] were chosen for the exchange-correlation functionals. The plane-wave cutoff energy was set to be 320 eV and the convergence of the force on each atom was set to be less than 0.001 eV/Å. Both the interlayer separation for the boron sheet and the intertubular distance for boron nanotubes were set to be 10 Å, which is enough to avoid any interactions between the periodic images. The Monkhorst-Pack scheme was used to sample the Brillouin zone.[26] Optimizations of both the cell parameters and atomic positions were carried out by minimization of the total energy.

## *Acknowledgements*

This work was supported by the National Basic Research Program of China (Grant No. 2006CB605101), the National Natural Science Foundation of China (Grant No.10874039) and the Natural Science Foundation of Hebei Province (No.A2008000134).

**Keywords:** Boron nanotube · Sparse boron sheet · Density functional theory

*Dr. J. Wang, Prof. Y. Liu and Prof. Y. C. Li*
*Department of Physics, and Hebei Advanced Thin Film Laboratory,*
*Hebei Normal University,*
*Shijiazhuang 050016, Hebei, China*
*National Key Laboratory for Materials Simulation and Design,*
*Beijing 100083, China*
*Fax: (+86) 311-86268314*
*\*Corresponding author E-mail: yliu@hebtu.edu.cn*

**Entry for the Table of Contents**

# ARTICLES

A new class of boron sheet and boron nanotube was predicted based on *ab initio* calculations. After relaxation the boron sheet remains flat and metallic. The nanotubes rolled from it, in our research scope, are metals except for the (8, 0) thin tube. Our predictions may be a good candidate for fabrication.

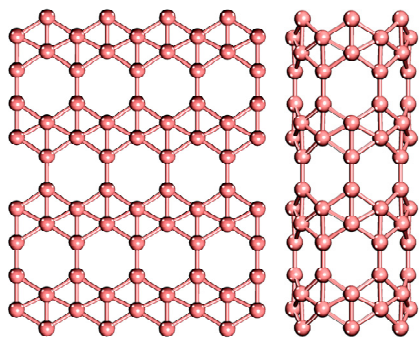

*Jing Wang, Ying Liu,\* and You-Cheng Li*

***Page No. – Page No.***

**A New Class of Boron Nanotube**